\documentstyle[prl,aps,twocolumn,psfig,harvard]{revtex}
\newcommand{\ea}{{\em et al.}}
\begin{document}
\draft
\title{Physical Aspects of Axonemal Beating and Swimming} 

\author{S\'ebastien Camalet and Frank J\"ulicher}
\address{PhysicoChimie Curie, UMR CNRS/IC 168, 26 rue d'Ulm, 75248 
Paris Cedex 05, France}
\maketitle

\begin{abstract}
We discuss a two-dimensional model for the dynamics of axonemal
deformations driven by internally generated forces of molecular
motors. Our model consists of an elastic filament pair connected by
active elements. We derive the dynamic equations for this system in
presence of internal forces. In the limit of small deformations, a
perturbative approach allows us to calculate filament shapes and the
tension profile. We demonstrate that periodic filament motion can be
generated via a self-organization of elastic filaments and molecular
motors. Oscillatory motion and the propagation of bending waves can
occur for an initially non-moving state via an instability termed Hopf
bifurcation. Close to this instability, the behavior of the system is
shown to be independent of microscopic details of the axoneme and the
force-generating mechanism. The oscillation frequency however does
depend on properties of the molecular motors. We calculate the
oscillation frequency at the bifurcation point and show that a large
frequency range is accessible by varying the axonemal length between 1
and 50$\mu$m. We calculate the velocity of swimming of a flagellum and
discuss the effects of boundary conditions and externally applied
forces on the axonemal oscillations.
\end{abstract}

\section{Introduction}

Many small organisms and cells swim in a viscous environment using
the active motion of cilia and flagella. These are hair-like
appendages of the cell which can undergo periodic motion and use
hydrodynamic friction to induce cellular self-propulsion
\cite{bray92}.  In this paper, we are interested in those flagella and
cilia which contain force generating elements integrated along the
whole length of the elastic filamentous structure. They represent
rod-like elastic structures which move and bend as a result of
internal stresses. Examples for these systems are paramecium which has
a large number of cilia on its surface; sperm, which use a single
flagellum to swim; and chlamydomonas which uses two flagella to swim
\cite{bray92}. Cilia also occur in very different situations. An
example is the kinocilium which exists in many hair bundles of
mechanosensitive cells and has the ability to beat periodically
\cite{rusc90}.

\begin{figure}
\centerline{\psfig{figure=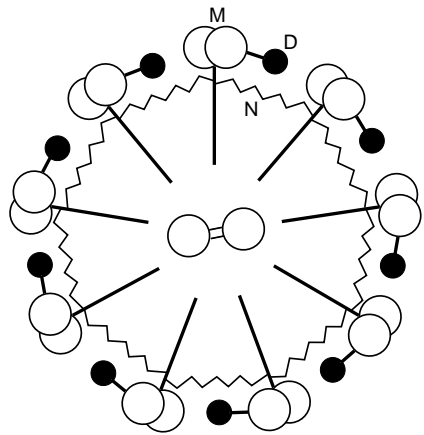,width=5.5cm}}
\bigskip
\caption{Schematic representation of the cross-section of the
axoneme. Nine doublets of microtubules (M) are arranged in a
cylindrical fashion, two microtubules are located in the
center. Dynein motors (D) are attached to microtubule doublets and
interact with a neighboring doublet. Elastic elements (N) such as
nexin are indicated as springs.}
\label{f:axo}
\end{figure}

The common structural theme of cilia and flagella is the axoneme, a
characteristic structure which occurs in a large number of very
different organisms and cells and which appeared early in
evolution. The axoneme consists of a cylindrical arrangement of 9
doublets of parallel microtubules and one pair of microtubules in the
center. In addition, it contains a large number of other proteins such
as nexin which provide elastic links between microtubule doublets, see
Fig. \ref{f:axo}.  The axoneme is inherently active. A large number of
dynein molecular motors are located in two rows between neighboring
microtubules and can induce forces and local displacements between
adjacent microtubules
\cite{albe94}.

Axonemal flagella can generate periodic waving or beating patterns of
motion. In the case of sperm for example, a bending wave of the
flagellum propagates from the head which contains the chromosomes
towards the tail. In a viscous environment, the surrounding fluid is
set in motion and hydrodynamic forces act on the filament. For typical
values of frequencies and length scales given by the size of a
flagellum, the Reynolds number is small and inertia terms in the fluid
hydrodynamics can be neglected \cite{tayl51}. Therefore, only friction
forces resulting from solvent viscosity can contribute to
propulsion. Under such conditions self-propulsion is possible if a
wave propagates towards one end, a situation which breaks
time-reversal invariance of the sequence of deformations of the
flagellum \cite{purc77}.

How can bending waves be generated by axonemal dyneins which act
internally within the axoneme? Dynein molecular motors induce relative
forces between parallel elastic filaments, which are microtubule
doublets.  As a result of these forces, the filaments have the
tendency to slide with respect to each other. If such a sliding is
permitted globally, the filaments simply separate but no bending
occurs. Bending results if global sliding is suppressed by rigidly
connecting the filament pair in the region close to one of the two
ends. In this situation, sliding is still possible locally, however
only if the filaments undergo a bending deformation. This coupling of
axonemal bending to local microtubule sliding has been demonstrated
experimentally.  In situations where the axoneme is cut at its basal
end, filaments slide and in the presence of ATP separate without
bending \cite{warn81}. Small gold-beads specifically attached to
microtubules in fully functioning flagella can be used to directly
visualize the local relative sliding during beating \cite{brok91}.

The theoretical problem of oscillatory axonemal bending and wave
patterns have been addressed by several authors.  One can distinguish
two principally different mechanisms to generate oscillatory forces
within the axoneme: (i) Deterministic forcing: a chemical oscillator
could regulate the dynein motors which are activated and deactivated
periodically.  In this case, the regulatory system defines a dynamical
force-pattern along the axoneme and drives the system in a
deterministic way \cite{sugi82}. (ii) Self-organized beating: the
axoneme oscillates spontaneously as a result of the interplay of
force-generating elements and the elastic filaments
\cite{mach63,brok75,brok85,lind95}.  In particular,  Brokaw has studied
thoroughly self-organized patterns of beating using numerical
simulations of simple models \cite{brok85,brok99}.  Such models are
based on the bending elasticity of the flagellum and an assumption on
the coupling of motor activity to the flagellar deformations.

In the present work, we are mostly interested in self-organized
beating. Our approach is conceptually different from other works as we
focus on an oscillating instability of the motor-filament system.  In
general, spontaneous oscillations occur via a so-called
Hopf-bifurcation where an initially stable quiescent state becomes
unstable and starts to oscillate. There is evidence for the existence
of such a dynamic instability in the axoneme.  Demembranated flagella
show a behavior which depends on the ATP-concentration $C_{ATP}$ in
the solution. For small $C_{\rm ATP}$, the flagellum is straight and
not moving. If $C_{\rm ATP}$ is increased, oscillatory motion sets in
at a critical value of the ATP concentration and persists for larger
concentrations \cite{gibb75}. This implies an instability of the
initial straight state with respect to a wave-like mode.  Recently, it
has been demonstrated using a simple model for molecular motors that a
large number of motors working against an elastic element can generate
oscillations via a Hopf bifurcation by a generic mechanism
\cite{juli95,juli97}.  This suggests that a system consisting only of
molecular motors and semiflexible filaments can in general undergo
self-organized oscillations. This idea is supported by the facts that
flagellar dyneins are capable of generating oscillatory motion
\cite{shin98} and that experiments suggest the existence of dynamic
transitions in many-motor systems \cite{rive98,fuji98}.  Patterns of
motion of cilia and flagella can be complex and embedded in three
dimensional space. Many examples of propagating bending waves,
however, are planar and can thus be considered as confined to
two-dimensions \cite{brok91}.

In the following sections, we present a systematic study of a
simplified model for the axoneme which has been introduced recently
\cite{cama99} and which captures the basic physical
properties that are relevant for its dynamics.  In Section II, we
present a thorough analysis of the dynamics of flexible filaments
driven by internal forces. This work is inspired by recent studies of
the dynamics of semiflexible filaments subject to external forces
\cite{gold95,wigg98,wigg98b}. We show how the
dynamic equations can be solved perturbatively and we calculate the
shapes of bending waves, the velocity of swimming and the tension
profile along the flagellum. In Section III, we briefly review a
simple two-state model for a large number of coupled molecular
motors. This model is well suited to represent the dynein molecular
motors which act within the axoneme.  Self-organized bending waves and
oscillations via a Hopf bifurcation of the coupled motor-filament
system are studied in section IV. We calculate the wave-patterns close
to a Hopf bifurcation and determine the frequencies selected by the
system. Finally, we discuss the relevance of our simple model to real
axonemal cilia and flagella and propose experiments which could be
performed to test predictions that follow from our work.

\section{A simple model for axonemal dynamics}

\begin{figure}
\centerline{\psfig{figure=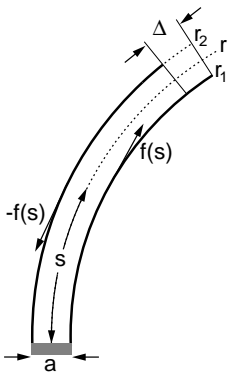,width=6.0cm}}
\bigskip
\caption{Two filaments (solid lines) ${\bf r_1}$ and ${\bf r_2}$
at constant separation $a$ are rigidly connected at the bottom end
with $s=0$, where $s$ is the arclength of the neutral line ${\bf r}$
(dashed).  Internal forces $f(s)$ are exerted in opposite directions,
tangential to the filaments. The sliding displacement $\Delta$ at the
tail is indicated.}
\label{f:filpair}
\end{figure}

The cylindrical arrangement of microtubule doublets within the axoneme
can be modeled effectively as an elastic rod. Deformations of this rod
lead to local sliding displacements of neighboring microtubules. Here,
we consider planar deformations. In this case, the geometric coupling
of bending and sliding can be captured by considering two elastic
filaments (corresponding to two microtubule doublets) arranged in
parallel with constant separation $a$ along the whole length of the
rod. At one end, which corresponds to the basal end of an axoneme and
which we call ``head'', the two filaments are rigidly attached and not
permitted to slide with respect to each other. Everywhere else,
sliding is possible, see Fig. \ref{f:filpair}.  The configurations of
the system are described by the shape of the filament pair given by
the position of the neutral line ${\bf r}(s)$ as a function of the
arclength $s$, where ${\bf r}$ is a point in two dimensional space.
The shapes of the two filaments are then given by
\begin{eqnarray}
{{\bf r}}_1(s) &=&{{\bf r}}(s) -a\;{\bf n}(s)/2 \nonumber \\
{{\bf r}}_2(s) &=&{{\bf r}}(s) +a\;{\bf n}(s)/2 \quad , \label{eq:r12}
\end{eqnarray}
where ${\bf n}$ with ${\bf n}^2=1$ is the filament normal.  The local
geometry along the filament pair is characterized by the relations
\begin{eqnarray}
\dot {\bf r} & = & {\bf t} \label{eq:dotr}\\
\dot {\bf t} & = & C {\bf n}\label{eq:dott}\\
\dot {\bf n} & = & -C {\bf t}\quad ,\label{eq:dotn}
\end{eqnarray}
where ${\bf t}$ denotes the normalized tangent vector and $C =
\dot{\bf t} \cdot {\bf n}$ is the local curvature.  Throughout this
paper dots denote derivatives with respect to $s$, i.e. $\dot{\bf
r}\equiv \partial_s {\bf r}\equiv \partial {\bf r}/\partial s$.

\subsection{Bending and sliding of a filament pair}

The two filaments are assumed to be incompressible and rigidly
attached to each other at one end where $s=0$. Bending of the filament
pair and local sliding displacements are then coupled by a geometric
constraint. The sliding displacement at position $s$ along the neutral
line
\begin{equation}
\Delta(s) \equiv \int_0^s ds' (|\dot {\bf r}_1|-|\dot {\bf r}_2|)
\label{eq:Deltadef}
\end{equation}
is defined to be the difference of the total arclengths along the two
filaments up to the points ${\bf r}_1(s)$ and ${\bf r}_2(s)$ which
face each other along the neutral line. From Eqns. (\ref{eq:r12}) and
(\ref{eq:dotn}) follows that $\dot {\bf r}_{1,2} = (1 \pm a C/2){\bf
t}$ and thus
\begin{equation}
|\dot {\bf r}_{1,2}| = 1 \pm a C/2 \quad .
\end{equation}
Therefore,
\begin{equation}
\Delta =  a\int_0^s ds' C \quad 
\label{eq:Delta}
\end{equation}
is given by the integrated curvature along the filament.

\subsection {Enthalpy functional}

The bending elasticity of filaments (microtubules) characterizes the
energetics of the filament pair.  In addition to filament bending, we
also have to take into account the large number of passive and active
elements (e.g. nexin links and dynein molecular motors) which give
rise to relative forces between neighboring microtubule
doublets. These internal forces can be characterized by a
coarse-grained description defining the force per unit length $f(s)$
acting at position $s$ in opposite directions on the two microtubules,
see Fig. \ref{f:filpair} (internal forces must balance). This force
density is assumed to arise as the sum of active and passive forces
generated by a large number of proteins.  This internal force density
corresponds to a shear stress within the flagellum which tends to
slide the two filaments with respect to each other.  A static
configuration of a filament pair of length $L$ can thus be
characterized by the enthalpy functional
\begin{equation}
G\equiv\int_0^{L}  \left[ \frac{\kappa}{2} C^2 + f \Delta + 
\frac{\Lambda}{2} \dot {\bf r}^2 \right ] ds \quad .
\label{eq:G}
\end{equation} 
Here, $\kappa$ denotes the total bending rigidity of the
filaments. The incompressibility of the system is taken into account
by the Lagrange multiplier function $\Lambda(s)$ which is used to
enforce the constraint $\dot{\bf r}^2=1$ which ensures that $s$ is the
arclength \cite{gold95}.  The internal force density $f$ couples to
the sliding displacement $\Delta$ as described by the contribution
$\int ds f \Delta$. The variation of this term under small
deformations of the filament shape represents the work performed by
internal stresses. Using Eq. (\ref{eq:Delta}) leads after a partial
integration to
\begin{equation}
G=\int_0^L ds\left[\frac{\kappa}{2} C^2 - a C F +
\frac{\Lambda}{2}\dot {\bf r}^2\right] \quad ,
\label{eq:GF}
\end{equation} 
where 
\begin{equation}
F(s) \equiv -\int_s^L  ds' f \quad 
\end{equation}
is the force density integrated to the tail. From Eq.(\ref{eq:GF}) it
follows that if the internal stresses or $F$ are imposed, $G$ is
minimized for a filament curvature $C=C_0$ where $C_0(s)=a F(s)/\kappa$
is a local spontaneous curvature. The internal forces therefore induce
filament bending.  In order to derive the filament dynamics, we
determine the variation $\delta G$ with respect to variations $\delta
{\bf r}$.  Details of this calculation are given in Appendix A. As a
result, we find
\begin{equation}
\frac{\delta G}{\delta {\bf r}} = 
\partial_s[\,(\kappa \dot C-a f)\,{\bf n} -\tau {\bf t}\,\,] 
\label{eq:dGdr} \quad .
\end{equation}
Here, 
\begin{equation}
\tau=\Lambda + \kappa C^2 - a C F \label{eq:tau0}\quad ,
\end{equation}
plays the role of the physical tension. This becomes apparent since 
from Eq. (\ref{eq:dGdr}) it follows that 
\begin{equation}
\tau(s)={\bf
t(s)}\cdot \left(\int_s^L ds' \delta G/\delta {\bf r} + {\bf F}_{\em ext}(L)\right)
\end{equation}
where ${\bf F}_{\em ext}(L)$ is the external force applied at the end
which satisfies Eq. (\ref{eq:bcle}). Therefore, $\tau$ is the tangent
component of the integrated forces acting on the filament.

\subsection{Dynamic equations}

We derive the dynamic equation with the simplifying assumption that
the hydrodynamics of the surrounding fluid can be described by two
local friction coefficients $\xi_\parallel$ and $\xi_\perp$ for
tangential and normal motion, respectively.  The
equations of motion in this case are given by  \cite{wigg98,wigg98b}
\begin{equation}
\partial_t {\bf r} = -\left(\frac{1}{\xi_{\perp}} {\bf n} {\bf n} +
 \frac{1}{\xi_{\parallel}} {\bf t} {\bf t}\,\right) \cdot
 {\frac{\delta G}{\delta {\bf r}}} \label{eq:gem1}\quad .
\end{equation}
For the following, it is useful to introduce a coordinate system ${\bf
r}=(X,Y)$ and the angle $\psi$ between the tangent ${\bf
t}=(\cos\psi,\sin\psi)$ and the X-axis which satisfies $C=\dot\psi$.  
We find with Eq. (\ref{eq:dGdr})
\begin{eqnarray}
 \partial_t {\bf r} &=& \frac{1}{\xi_{\perp}} {\bf n}\, (-\kappa
 \stackrel{\mbox{...}}{\psi} + a \dot f + \dot \psi \tau) 
\nonumber \\ &+&
 \frac{1}{\xi_{\parallel}} {\bf t} \, (\kappa \dot \psi \ddot \psi
 -a\dot \psi f + \dot \tau) \quad .
\label{eq:gem2}
\end{eqnarray}  
Noting that $\partial_t\dot{\bf r}={\bf n} \partial_t \psi$, we obtain
an equation of motion for $\psi(s)$ alone:
\begin{eqnarray}
\partial_t \psi &=& \frac{1}{\xi_{\perp}}
(-\kappa\stackrel{\mbox{....}}{\psi} + a \ddot f + \dot \psi \dot \tau
+\tau \ddot \psi ) \nonumber \\
&+& \frac{1}{\xi_{\parallel}} \dot \psi (\kappa \dot
\psi \ddot \psi -af\dot \psi + \dot \tau) \quad .
\label{eq:emt}
\end{eqnarray}  

The tension $\tau$ is determined by the
constraint of incompressibility $\partial_t \dot{\bf r}^2= 2 {\bf t} 
\cdot \partial_t \dot{\bf r}=0$.  This
condition and Eq. (\ref{eq:gem2}) leads to a differential equation for
the tension profile:
\begin{equation}
\ddot \tau - \frac{\xi_{\parallel}}{\xi_{\perp}}\dot \psi^2 \tau = a
\partial_s(\dot \psi f)-\kappa \partial_s (\dot \psi \ddot \psi) +
\frac{\xi_{\parallel}}{\xi_{\perp}} \dot \psi (a\dot f - \kappa
\stackrel{\mbox{...}}{\psi}) \quad .
\label{eq:li}
\end{equation}
Eqns. (\ref{eq:emt}) and (\ref{eq:li}) determine the filament
dynamics.  The filament shape follows from
\begin{equation}
{\bf r}(s,t)= {\bf r}(0,t)+ \int_0^s (\cos \psi,\sin \psi) ds'  \quad ,
\label{eq:sdfromtv}
\end{equation}
where ${\bf r}(0,t)$ can be obtained from Eq. (\ref{eq:gem2})
evaluated at $s=0$.

\subsection{Boundary Conditions}

The filament dynamics depends on the imposed boundary conditions.  The
variation $\delta G$ has contributions at the boundaries which can be
interpreted as externally applied forces ${\bf F}_{\rm ext}$ and
torques $T_{\rm ext}$ acting at the ends, see Appendix A.  At the head
with $s=0$,
\begin{eqnarray}
{\bf F}_{\rm ext}&=&(\kappa \dot C-af){\bf n}-\tau {\bf t}\nonumber \\
T_{\rm ext}&=& -\kappa C-a\int_0^Lds' f\quad \label{eq:bcla}.
\end{eqnarray}
Similarly, at the tail for $s=L$,
\begin{eqnarray}
{\bf F}_{\rm ext}&=&(-\kappa \dot C+af){\bf n}+\tau {\bf t} \nonumber\\
T_{\rm ext}&=& \kappa C\quad \label{eq:bcle} .
\end{eqnarray}
If constraints on the positions and/or angles are imposed at the ends,
forces and torques have to be applied to satisfy these constraints.
In the following, we will discuss different boundary conditions as
specified in Table \ref{t:bc}:

Case A is the situation of a filament with clamped head, i.e. both the
tangent and the position at the head are fixed, the tail is free.
Case B is a filament with fixed head, i.e. the tangent at the head can
vary. The situation of a swimming sperm corresponds to case C where
the friction force of a viscous load attached at the head is taken
into account, the head is otherwise free.  As an example of a
situation with an external force ${\bf F}_{\rm ext}(L)$ applied at the
tail we consider Case D. For simplicity, we assume a force parallel to
the (fixed) tangent at the head.

\begin{table}
\begin{tabular}{|c|lr|lr|}
boundary &     head $s=0$  &  &     tail  $s=L$  &\\
condition&             &  &                  & \\
\hline
A &  $\partial_t{\bf r}={\bf 0}$ & $\partial_t {\bf t}={\bf 0}$ &  
                        ${\bf F}_{\rm ext}={\bf 0}$ & $T_{\rm ext}=0$ \\
B &  $\partial_t {\bf r}={\bf 0}$ & $T_{\rm ext}=0$ &
                        ${\bf F}_{\rm ext}={\bf 0}$ & $T_{\rm ext}=0$ \\
C &  ${\bf F}_{\rm ext}=-\zeta \partial_t {\bf r}$ & $T_{\rm ext}=0$ &
                        ${\bf F}_{\rm ext}={\bf 0}$ & $T_{\rm ext}=0$  \\
D &  $\partial_t {\bf r}={\bf 0}$ & $\partial_t {\bf t}={\bf 0}$ &
                        ${\bf F}_{\rm ext} \neq {\bf 0}$ & $T_{\rm ext}=0$ \\
\end{tabular}
\caption{Different boundary conditions studied. (A) clamped head, free tail;
(B) fixed head, free tail; (C) swimming flagellum with viscous load
$\zeta$; (D) clamped head, external force applied at the tail.}
\label{t:bc}
\end{table}

\subsection{Small deformations}

In the absence of internal forces $f$, the filament relaxes passively
to a straight rod with $\dot \psi=0$, i.e. $\psi$ constant.  Without
loss of generality, we choose $\psi=0$ in this state which implies
that straight filament is parallel to the $X$-axis.  Internal stresses
$f(s)$ induce deformations of this straight conformation. For small
internal stresses, we can perform a systematic expansion of the
filament dynamics in powers of the stress amplitude.

\begin{figure}
\centerline{\psfig{figure=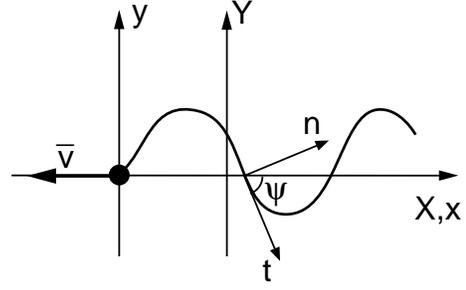,width=6.0cm}}
\bigskip
\caption{Resting frame $(X,Y)$ and frame $(x,y)$
moving with the filament at velocity $\bar v$. The tangent ${\bf t}$
and the normal ${\bf n}$ are indicated, the angle between the
$X,x$-axis and ${\bf t}$ is denoted $\psi$.}
\label{f:swim}
\end{figure}

We introduce a dimensionless parameter $\epsilon$ which scales the
amplitudes of the internal stresses, $f(s,t)=\epsilon f_1(s,t)$, where
$f_1$ is an arbitrary stress distribution. We can now solve the
dynamic equations (\ref{eq:emt}) and (\ref{eq:li}) perturbatively by
writing
\begin{eqnarray}
\psi &=&\epsilon \psi_1+\epsilon^2 \psi_2+O(\epsilon^3) \nonumber \\
\tau &=& \tau_0+\epsilon \tau_1+\epsilon^2 \tau_2+O(\epsilon^3) 
\label{eq:exqpsi}  \quad ,
\end{eqnarray}
which allows us to determine the coefficients $\psi_n(s,t)$ and
$\tau_n(s,t)$.  This procedure is described in Appendix B.  We find
that $\psi_2$ and $\tau_1$ always vanish and $\tau_0=\sigma$ is a
constant tension which is equal to the $X$-component of the external
force applied at the end.  Note, that for boundary conditions A-C,
$\sigma=0$. The filament shape is thus characterized by the behavior
of $\psi_1(s,t)$ which obeys
\begin{equation}
\xi_{\perp} \partial_t \psi_1 = - \kappa \stackrel{\mbox{....}} {\psi_1} + 
\sigma \ddot {\psi_1} + a \ddot {f_1} \quad .
\label{eq:emt2}
\end{equation} 
In order to discuss the filament
motion in space, we define the average velocity of swimming

\begin{equation}
\bar {\bf v} = \lim_{t \rightarrow \infty} \frac{1}{t} \int_0^t dt' \partial_t {\bf r}
\end{equation}
which is independent of $s$. This velocity is different from zero only
in the case of boundary condition C. We choose the $X$-axis parallel
to $\bar{\bf v}$ and introduce a
coordinate system $(x,y)$ which moves with the filament
\begin{equation}
(x,y)=(X-\bar v t,Y) \quad , \label{eq:movfr}
\end{equation}
where $\bar v=\pm|\bar {\bf v}|$ and the sign depends on whether
motion is towards the positive or negative $X$-direction, see
Fig. \ref{f:swim}. It is useful to introduce the transverse and
longitudinal deformations $h$ and $u$, respectively, which satisfy
\begin{equation}
(x,y) = \left( s + u(s) , h(s) \right) \label{eq:uh}
\end{equation}
and which vanish for $\epsilon=0$. The quantities $h$, $u$ and $\bar
v$ can be calculated perturbatively in $\epsilon$, see Appendix B.
To second order in $\epsilon$ transverse motion satisfies
\begin{equation}
\xi_{\perp} \partial_t h= - \kappa \partial_s^4 h + \sigma \partial_s^2 h + 
a \partial_s f 
\label{eq:emld}
\end{equation}  Longitudinal displacements satisfy
\begin{equation}
u(s)=u(0)-\frac{1}{2}\int_0^s (\partial_s h)^2 ds' 
\label{eq:u2x0}
\end{equation}
The boundary conditions for $h(s)$ are given in table
\ref{t:mongebc}.

\subsection{Swimming} 

The velocity of swimming $\bar v$ can be calculated perturbatively in
$\epsilon$. We consider for simplicity periodic internal stresses with
frequency $\omega$ given by\footnote{Note, that more general periodic
stresses $f=\sum_{n}\tilde f_n e^{in\omega t}$ can also be
considered. Since the equation of motion (\ref{eq:emld}) is linear,
different modes superimpose linearly and we can without loss of
generality restrict our discussion to a single mode $n=1$.}
\begin{equation}
f(s,t)=\tilde f(s)^{i\omega t}+\tilde {f^{*}}(s) e^{-i\omega t} \quad
\end{equation}
which after long time leads to periodic filament motion
\begin{equation}
h(s,t)=\tilde h(s) e^{i\omega t}+\tilde {h^{*}}(s) e^{-i\omega t}
\end{equation}
where $\tilde h$ satisfies according to Eq. (\ref{eq:emld})
\begin{equation}
\kappa \partial_s^4\tilde h - \sigma \partial_s^2 
\tilde h+\xi_{\perp} i \omega \tilde h=a \partial_s \tilde f \quad .
\label{eq:fc}
\end{equation}

\begin{table}
\begin{tabular}{|c|lr|lr|}
boundary &     head $s=0$  &  &     tail  $s=L$  & \\
condition&       $F=-\int_0^L ds f$      &  &      $\partial_s^2 h=0$  & \\
\hline
A &  $h=0        $ & $\partial_s h=0$ &  
                          $\kappa\partial_s^3 h = a f$ \\
B &  $h=0        $ & $\kappa \partial_s^2 h=aF$ &
                          $\kappa\partial_s^3 h = a f$ \\
C &  $\zeta\partial_t h=af-\kappa \partial_s^3 h $ & $\kappa \partial_s^2 h =aF$ &
                         $\kappa\partial_s^3 h = a f$ \\
D &  $h=0$ & $\partial_s h=0$ &
                         $\kappa\partial_s^3 h $ \\
 &   &  &
                         $ -\sigma \partial_s h = a f$ \\
\end{tabular}
\caption{Boundary conditions for small amplitude motion.  
(A) clamped head, free tail; (B) fixed head, free tail; (C) swimming
flagellum with viscous load $\zeta$; (D) clamped head, external force
 applied at the tail.}
\label{t:mongebc}
\end{table}

\begin{figure}
\centerline{\psfig{figure=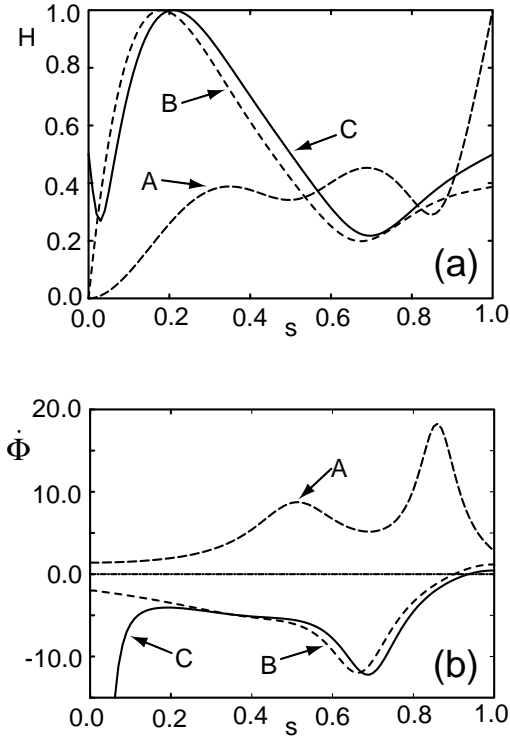,width=7.0cm}}
\bigskip
\caption{ 
Fundamental~modes~of~filament motion $\tilde h_1(s)=H e^{-i\phi}$ close
to a Hopf bifurcation are displayed for an oscillation frequency of
about $50 Hz$ and boundary conditions A-C.  In case C, a viscous load
$\zeta \simeq 5.10^{-8} Ns / m$ acts at the head. (a) Amplitude $H$ in
arbitrary units as a function of arclength $s$. (b) Same plot for the
gradient of the phase $\dot\phi$.}
\label{f:HPhi}
\end{figure}

\begin{figure}
\centerline{\psfig{figure=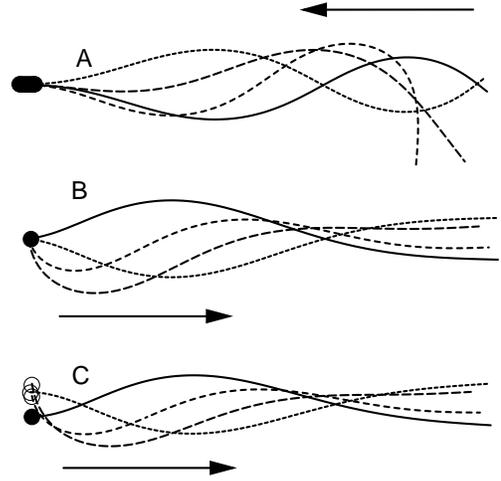,width=6.5cm}}
\bigskip
\caption{Filament shapes in the $(x,y)$ plane that correspond to the
modes displayed in Fig. \protect\ref{f:HPhi}. Snapshots taken at
different times to illustrate bending waves are shown for boundary
conditions A-C. The arrows indicate the direction of wave
propagation.}
\label{f:beat}
\end{figure}

Writing $\tilde h=H e^{-i\phi}$ where $H$ and $\phi$ denote the
amplitude and phase, respectively, filament motion can be
expressed as
\begin{equation}
h(s,t)=H(s) \cos(\omega t-\phi(s)) \quad ,
\end{equation}
which represents bending waves for which $v_p=\omega/\dot\phi$ can be
interpreted as the local wave propagation velocity.  Examples of
bending waves for self-organized beating discussed in section IV are
displayed in Figs. \ref{f:HPhi} and \ref{f:beat}.  For a freely
oscillating filament with a viscous load of friction coefficient
$\zeta$ attached (case C), we find an average propulsion velocity
$\bar v=V_0/(1+\zeta/\xi_{\parallel}L)$ where
\begin{equation} V_0 = -
\left(\frac{\xi_{\perp}}{\xi_{\parallel}}-1\right)
\frac{\omega}{2 L} \int_0^L ds H^2 \dot \phi \quad ,
\label{eq:noloadavve}
\end{equation}

is the velocity for $\zeta=0$. Eq. (\ref{eq:noloadavve}) which is
correct to second order in $\epsilon$ reveals that motion is only
possible if the filament friction is anisotropic
$\xi_{\parallel}/\xi_{\perp}\neq 1$ and if a wave is propagating,
$\dot \phi\neq 0$. This is consistent with earlier work on swimming
\cite{tayl51,purc77,ston96}.  For a rod-like filament with
$\xi_{\parallel} < \xi_{\perp}$, swimming motion is opposite to the
direction of wave propagation.

\section{Internal forces generated by molecular motors}

The filament dynamics is driven by internal stresses $f(s,t)$
generated by a large number of dynein motors and passive elements. We
describe the properties of a force-generating system using a simplified
two-state model.

\subsection{Two state model for molecular motors}

We briefly review the two-state model for a large number of motor
molecules attached to a filament which slides with respect to a second
filament \cite{juli95,juli97b}.  Each motor is assumed to have two
different chemical states, a strongly bound state $1$ and a weakly
bound state or detached state $2$. The interaction between a motor and
a filament in states $1$ and $2$ is characterized by energy landscapes
$W_1(x)$ and $W_2(x)$, where $x$ denotes the position of a motor along
the filament.  The potentials reflect the symmetry of the filaments:
they are periodic with period $l$, $W_i(x)=W_i(x+l)$ and are spatially
asymmetric, $W_i(x)\neq W_i(-x)$. In the presence of ATP which is the
chemical fuel that drives the system, the motors are assumed to
undergo transitions between states. The corresponding transition rates
are denoted $\omega_1$ for detachments and $\omega_2$ for attachments.
We introduce the relative position $\xi$ of a motor with respect to
the potential period where $x=\xi+ nl$, $0 \le
\xi <l$ and $n$ is an integer. The probability to find a
motor in state $i$ at position $\xi$ at time $t$ is denoted
$P_i(\xi,t)$, $P_1+P_2$ is normalized within one period.  The dynamic
equations of the system are given by
\begin{eqnarray}
\partial_t P_1+v \partial_{\xi} P_1 &=& -\omega_1 P_1 + \omega_2 P_2 \nonumber \\
\partial_t P_2+v \partial_{\xi} P_2 &=& \omega_1 P_1 - \omega_2  P_2  \quad .
\end{eqnarray}
The sliding velocity $v=\partial_t x$ is determined by the force-balance
\begin{equation}
f= \lambda v +\rho \int_0^l (
P_1\partial_{\xi}W_1+P_2\partial_{\xi}W_2) d\xi \ + K x \quad
. \label{eq:fdens}
\end{equation}
Here, the coefficient $\lambda$ describes the total friction per unit
length in the system. The number density of motors along the filament
is denoted by $\rho$ and $f$ is the force per unit length generated by
the system. The elastic modulus per unit length $K$ occurs in presence
of elastic elements such as nexins.  If motors have an incommensurate
arrangement compared to the filament periodicity, $P_1+P_2=1/l$ and
the equations simplify and can be expressed by $P_1$ alone
\begin{eqnarray}
\partial_t P_1 &+& v \partial_\xi P_1 = -(\omega_1 + \omega_2 )P_1 + 
\omega_2/l\nonumber  \\
f & = &  \lambda v + \rho \int_0^l P_1 \partial_\xi \Delta W  + K x
\label{eq:twost}
\end{eqnarray}
where $\Delta W=W_1-W_2$. 

\subsection{Molecular motors coupled to a filament pair}

The two state model for a large number of motors is well suited to
represent the internal forces acting within the axoneme.  We assume
that at any position $s$ an independent two-state model described by
Eq. (\ref{eq:twost}) is located which generates the internal force
density $f(s)$. The filament sliding displacement is identified with
the motor displacement, $\Delta\equiv x$.  Note, that here we neglect
for simplicity fluctuations which arise from the chemical activity of
a finite number of force generators and we assume homogeneity of all
properties along the axonemal length.  The energy source of the active
system is the chemical activity characterized by the transition rates
$\omega_1$ and $\omega_2$, the generated forces induce bending
deformations of the filaments.

The dynamic equations (\ref{eq:twost}) represent a nonlinear active
system which generates time-dependent forces. Since we will study
periodic motion, we can express forces and displacements by a Fourier
expansion
\begin{eqnarray}
f(t) & = & \sum_n f_n e^{i n \omega t}\\
\Delta(t) & = & \sum_n \Delta_n e^{i n \omega t}\quad .
\end{eqnarray}
The relation between force amplitudes $f_n$ and amplitudes of sliding
displacements $\Delta_n$ can in general be expressed in powers of the
amplitudes $\Delta_n$ \cite{juli97}
\begin{equation}
f_n = F^{(1)}_{nk} \Delta_k + F^{(2)}_{nkl} \Delta_k
\Delta_l + O(\Delta_k^3) \quad . \label{eq:fnexp}
\end{equation}
Here, the summation over common indices is implied. The coefficients
$F^{(n)}$ are calculated for the two-state model in
Appendix C. However, Eq. (\ref{eq:fnexp}) is more general and
characterizes a whole class of active nonlinear mechanical systems.

\subsection{Symmetry considerations}

The interaction of the motors with the two filaments is asymmetric.
Motors are rigidly connected to one of the filaments and slide along
the second.  In addition, the filaments are polar and filament sliding
is induced by motors towards one particular end of the two filaments.
As a consequence, the system has a natural tendency to create bending
deformations with one particular sign of the curvature. Exchanging the
role of the two filaments r everses this sign of bending
deformations. This broken symmetry does however not reflect the
symmetry of the axoneme which is symmetric with respect to all
microtubule doublets. Each of the nine microtubule doublet within the
cylindrical arrangement of the axoneme play identical roles and
interact with rigidly attached motors as well as with motors which
slide along their surfaces. If all motors generate a constant force,
there is consequently no resulting bending deformation since all
bending moments cancel by symmetry.

\begin{figure}
\centerline{\psfig{figure=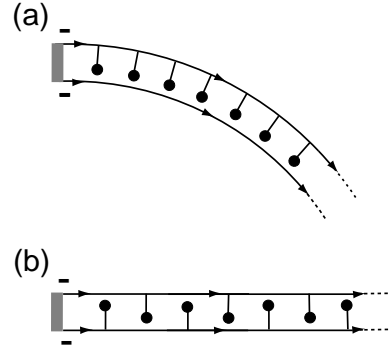,width=5.0cm}}
\bigskip
\caption{Asymmetric (a) and symmetric (b) motors/filament pair. The 
arrows indicate the polarity of the filaments. In case (a),
spontaneous bending occurs. In case (b) both filaments play identical
roles, no spontaneous bending occurs.}
\label{f:sym}
\end{figure}

 In order to introduce this
axonemal symmetry in our two-dimensional model, we assume that both
filaments have motors which are rigidly attached and which slide on
the second filament, see Fig. \ref{f:sym}. The consequence of this
symmetry is that the filament polarity becomes unimportant.
Exchanging the role of the two filaments is equivalent to replacing
the sliding displacement $\Delta$ of motors by $-\Delta$. Therefore,
in the context of the two-state model this symmetrization is
equivalent to the requirement that energy landscapes $W_i(x)$ and
transition rates $\omega_i(x)$ are symmetric functions with respect to
$x \rightarrow -x$.  As a consequence of this symmetry, there is no
preferred direction of bending. In the following, we adopt this choice
of a symmetric motor/filament pair.  If we use the expansion given in
Eq. (\ref{eq:fnexp}), this symmetry implies that all even coefficients
$F^{(2n)}=0$.

\subsection{Linear response function}

The perturbative treatment to study small filament deformations
introduced in section II can be naturally extended to the coupled
motor-filament situation using the expansion (\ref{eq:fnexp}). Up to
second order in $\epsilon$, only the linear response coefficient of
the active system plays a role which for the two-state model is given
by $F^{(1)}_{nk} = \chi \delta_{nk}$ with
\begin{equation}
\chi =K+  \lambda i\omega- \rho \int_0^l d\xi \, \partial_\xi \Delta W
\frac{\partial_\xi R 
}{\alpha+i\omega} i\omega \quad .
\label{eq:F1}
\end{equation}
Here, we have introduced $R \equiv \omega_2/\alpha l$ and
$\alpha=\omega_1+\omega_2$.  Higher order terms $F^{(2n+1)}$ have to
be taken into account if the third or higher order in $\epsilon$ is
considered.  The linear response function $\chi$ as well as nonlinear
coefficients can be calculated most easily for a simple choice of
symmetric potential and transition rates
\begin{eqnarray}
\Delta W(\xi) & = & U \cos (2\pi \xi/l) \nonumber \\
\omega_1(\xi) & = & \beta - \beta\cos(2 \pi \xi/l) \nonumber\\
\omega_2(\xi) & = & \alpha -\beta + \beta\cos(2 \pi\xi/l)
\end{eqnarray}
where $\alpha$ and $\beta$ are $\xi$-independent rate constants. For
this convenient choice
\begin{equation}
\chi(\Omega,\omega)= K + i \lambda \omega  - \rho k \Omega
 \frac{i\omega/\alpha+(\omega/\alpha)^2}{1+(\omega/\alpha)^2} \quad ,
\label{eq:linrep}
\end{equation}
where $k\equiv U/l^2$ is the cross-bridge elasticity of the motors.
We have introduced the dimensionless parameter $\Omega =
2\pi^2\beta/\alpha$ with $0<\Omega<\pi^2$ which plays the role of a
control parameter of the motor-filament system, $\alpha$ is a
characteristic ATP cycling rate. 

\section{Self-organized beating via a Hopf bifurcation}

A Hopf-bifurcation is an oscillating instability of an initial
non-oscillating state which occurs for a critical value $\Omega_c$ of
a control parameter. For $\Omega < \Omega_c$, the system is passive
and not moving, while for $\Omega>\Omega_c$ it exhibits spontaneous
oscillations.  As we demonstrate below, self-organized oscillations of
the driven filament pair are a natural consequence of its physical
properties. The control parameter $\Omega$ is in our two-state model a
ratio of chemical rates of the ATP hydrolysis cycle. In an
experimental situation, it could be varied e.g. by changing the ATP
concentration. If the molecular motors are regulated by some other ion
concentration such as e.g. $Ca^{2+}$, this concentration could also
play the role of the control parameter.

\subsection{Generic aspects}

For oscillations in the vicinity of a Hopf-bifurcation, filament
deformations are small and we can thus use Eq. (\ref{eq:emld}) to
describe the filament dynamics.  The force amplitude $\tilde f$ can be
obtained from the expansion
\begin{equation}
\tilde f=\chi(\Omega,\omega) \tilde \Delta + O(\tilde \Delta^3) 
\quad \label{eq:linres}
\end{equation}
in the amplitude of sliding displacements. In addition, this amplitude
is related to the filament shape:
\begin{equation}
\Delta(s) = a (\partial_s h(s) - \partial_s h(0)) + O(h^3) \label{eq:Deltalin} \quad .
\end{equation}
With Eq. (\ref{eq:fc}), we find that spontaneously oscillating modes
$\tilde h(s)$ at frequency $\omega$ are to linear order near the
bifurcation solutions to
\begin{equation}
\kappa \partial_s^4\tilde h - \sigma \partial_s^2 
\tilde h + \xi_{\perp} i \omega \tilde h = a^2 \chi \partial_s^2 \tilde h
\label{eq:ep}
\end{equation}
Note that this equation is general and its structure does not depend
on the specific model chosen for the force generating elements.
Boundary conditions corresponding to cases A-D follow from Table
\ref{t:mongebc} by replacing $h\rightarrow
\tilde h$, $\partial_t h \rightarrow i\omega \tilde h$ and $f
\rightarrow \tilde f= a\chi(\partial_s \tilde h -
\partial_s \tilde h(0))$.

As discussed in Appendix D, Eq. (\ref{eq:ep}) has the structure of an
eigenvalue problem, with $\chi$ playing the role of a complex
eigenvalue.  For every choice of parameters, there exists an infinite
set of nontrivial solutions $\tilde h_n$ to Eq. (\ref{eq:ep}) if
$\chi$ is equal to the corresponding eigenvalue $\chi_n$, $n=1,2,..$.
We order these values according to their amplitude:
$|\chi_n|\leq|\chi_{n+1}|$.  The functional dependence of the
Eigenvalues $\chi_n$ on the model parameters is completely determined
by dimensionless functions $\bar\chi_n(\bar\sigma,\bar\omega)$
according to
\begin{equation}
\chi_n = \frac{\kappa}{L^2 a^2} \bar \chi_n (\bar \sigma, \bar \omega) \quad ,
\end{equation}
where $\bar \sigma \equiv \sigma L^2 /\kappa$ and $\bar \omega \equiv
\omega \xi_{\perp} L^4/\kappa$ are a dimensionless tension and
frequency, respectively.

The existence of discrete eigenvalues reveals that only particular
values of the linear response function of the active material are
consistent with being in the vicinity of a Hopf-bifurcation.  If the
actual value of $\chi(\Omega,\omega)$ of the system differs from one
of the eigenvalues $\chi_n$, the system is either not oscillating, or
it oscillates with large amplitude for which higher order terms in
$\epsilon$ cannot be neglected.  An important consequence of this
observation is that the modes $\tilde h_n$ of filament beating close
to a Hopf-bifurcation can be calculated without knowledge of the
properties of the active elements. Not even knowledge of the linear
response coefficient is required since it follows as an eigenvalue at
the bifurcation.  Therefore, filament motion close to a Hopf
bifurcation has generic or universal features which do not depend on
details of the problem such as the structural complexity within the
axoneme. The motion is given by one of the modes $\tilde h_n$ which
only depend on $\bar \sigma$ and $\bar
\omega$.

\begin{figure}
\centerline{\psfig{figure=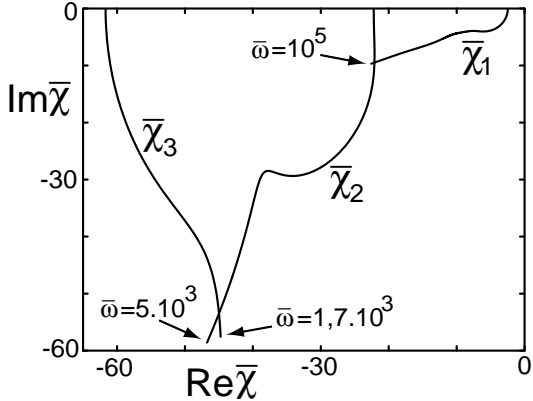,width=7.0cm}}
\bigskip
\caption{(a) Eigenvalues $\bar \chi_n$ for boundary conditions B.  Each
eigenvalue is represented by a line which traces the locations of
$\bar\chi_n$ in the complex plane for varying reduced frequency
$\bar\omega$. The lines begin for $\bar
\omega=0$ on the real axis, the value of $\bar \omega$ at the
other ends are indicated.}
\label{f:chi}
\end{figure}

Fig. \ref{f:HPhi} displays the amplitude $H$ and the gradient of the
phase $\dot \phi$ of the fundamental mode $\tilde h_1=H
e^{-i\phi}$. Note, that we use arbitrary units for $H$ since the
amplitude of $\tilde h(s)$ is not determined by Eq. (\ref{eq:ep}).
Corresponding filament shapes represented according to
Eq. ({\ref{eq:uh}) in the $(x,y)$ plane, are displayed in
Fig. \ref{f:beat} for boundary conditions A-C as snapshots taken at
different times. The oscillation amplitudes are chosen in such a way
that the maximal value of $\psi\simeq \pi/2$.  The direction of wave
propagation depends on the boundary conditions. For clamped head (case
A), waves propagate towards the head whereas for fixed or free head
(cases B and C), they propagate towards the tail.  The positions of
the lowest eigenvalues $\bar
\chi_n$ in the complex plane for boundary conditions B are displayed
in Fig. \ref{f:chi} for $\bar\sigma=0$ and varying $\bar
\omega$.

\subsection{Selection of eigenmodes and frequency}

In order to determine which of the modes $\tilde h_n$ is selected and
at what frequency $\omega$ the system oscillates at the bifurcation
point, explicit knowledge of the linear response function
$\chi(\Omega,\omega)$ is necessary.  As a simple example, we use
$\chi$ as given by Eq. (\ref{eq:linrep}) for a two-state model. For
$\Omega=0$, $\chi=K+i \lambda\omega$ which is a passive viscoelastic
response. In this case, no spontaneous motion is possible which can be
seen by the fact that all eigenvalues $\bar
\chi_n$ have negative real and imaginary parts which cannot be matched
by the linear response of a passive system. If $\Omega$ is increased,
we are interested in a critical point $\Omega=\Omega_c$ for which the
straight filament configuration becomes unstable. This happens as soon
as the linear response function $\chi$ matches for a particular
frequency $\omega=\omega_c$ one of the complex eigenvalues,
\begin{equation}
\frac{\kappa}{a^2 L^2} \bar \chi_n(\bar \sigma,\bar \omega_c)
=\chi(\Omega_c,\omega_c) \quad , \label{eq:crit}
\end{equation}
where $\bar\omega_c = \xi_\perp L^4\omega_c/\kappa$.  Since $\chi$ is
complex, both the real and the imaginary part of Eq. (\ref{eq:crit})
represent independent conditions.  Therefore, Eq. (\ref{eq:crit})
determines both the critical point $\Omega_c$ as well as the selected
frequency $\omega_c$. The selected mode $n$ is the first mode
(beginning from $\Omega=0$) to satisfy this condition. Since
$|\chi(\Omega,\omega)|$ typically increases with increasing $\Omega$,
the instability occurs almost exclusively for $n=1$ since $|\bar
\chi_1|$ has been defined to have the smallest value.
For $\Omega > \Omega_c$, but close to the transition, the system
oscillates spontaneously with frequency $\omega_c$. The shapes of
filament beating are characterized by the corresponding mode $\tilde
h_n(s)$ whose amplitude is determined by nonlinear terms. For the case
of a continuous transition, the deformation amplitude increases as
$|\tilde h_n|\sim (\Omega-\Omega_c)^{1/2}$.

\subsection{Axonemal vibrations for different lengths}

We choose the parameters of our model to correspond to the axonemal
structure.  We estimate the bending rigidity $\kappa\simeq 4\cdot
10^{-22}$Nm$^2$ which is the rigidity of about 20
microtubules. Furthermore, we choose a microtubule separation $a\simeq
20$nm, motor density $\rho\simeq 5\cdot 10^8$m$^{-1}$, friction per
unit length $\lambda\simeq 1 $kg/ms, rate constant $\alpha\simeq
10^3$s$^{-1}$, cross-bridge elasticity $k\simeq 10^{-3}$N/m and a
perpendicular friction $\xi_\perp\simeq 10^{-3}$kg/ms which is of the
order of the viscosity of water. A rough estimate of the elastic
modulus per unit length associated with filament sliding can be
obtained by comparing the number of dynein heads to the number of
nexin links within the axoneme. This suggests that $K$ is relatively
small, $K {<\atop \sim } k\rho/10$.

The selected frequency $\omega_c$ at the bifurcation point for case B
is shown in Fig. \ref{f:frL} as a function of the axoneme length for
$K=0$. For small lengths, the oscillation frequency is large and
increases as $L$ decreases. In this high-frequency regime, which
occurs for $L{<\atop
\sim} 10\mu$m, the system vibrates in a mode with no apparent wave
propagation.  For $L{>\atop \sim} 10\mu$m the frequency is only
weakly $L$-dependent and the system propagates bending waves of a
wave-length shorter than the filament length.

The limit of small lengths corresponds to small $\bar\omega$ and can
be studied analytically. For $\bar\sigma=0$, the eigenvalue $\chi_1$
is given to linear order in $\bar\omega$ by $\bar \chi_1 \simeq -
\pi^2 / 4 - i \gamma \bar \omega$. 
The coefficient $\gamma$ depends on boundary conditions but not on any
model parameters. For a clamped head (A), $\gamma\simeq 0.184$, for a
fixed head (B), $\gamma\simeq 0.008$, see Appendix D.  The criterion
for a Hopf bifurcation for small $L$ is
\begin{equation}
\chi (\Omega_c,\omega_c) \simeq  -\frac{\pi^2\kappa}{4 a^2 L^2} 
- i \gamma \frac{\xi_{\perp}\omega_c L^2}{a^2}
\label{eq:chi1approx}
\end{equation} 
Together with Eq.(\ref{eq:linrep}), we find the critical frequency
\begin{equation}
\omega_c \simeq \frac{\pi}{2 L^2}\left( \frac{\kappa \alpha}
{\gamma \xi_\perp}\right)^{1/2}
\left[\frac{1+(L/L_{K})^2}{1+(L_\lambda/L)^2}\right]^{1/2}
\label{eq:omegaapprox} \quad ,
\end{equation}
where $L_{K}\equiv(\pi^2 \kappa/4K a^2)^{1/2}$ and
$L_\lambda\equiv(\lambda a^2/\gamma\xi_\perp)^{1/2}$ are two
characteristic lengths. For $K$ and $\lambda$ small, $L_{K}\gg L \gg
L_\lambda$ and the critical frequency behaves as $\omega_c \sim
1/L^2$.

\centerline{\psfig{figure=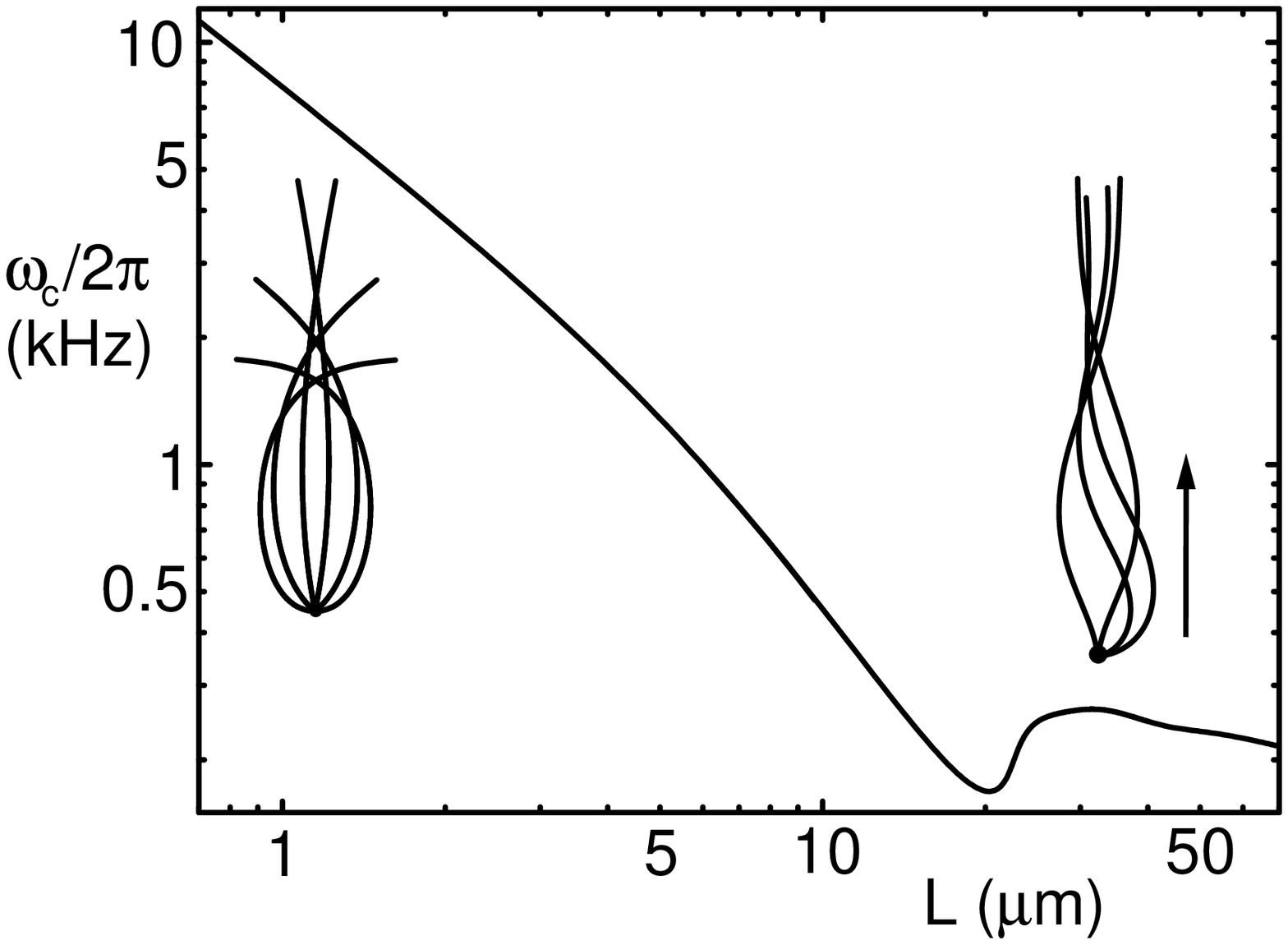,width=7.0cm}}
\bigskip
\begin{figure}
\caption{Oscillation frequency $\omega_c/2\pi$ at the bifurcation
point as a function of the axoneme length $L$ for boundary conditions
B and model parameters as given in the text. The insets show
characteristic patterns of motion for small and large lengths.}
\label{f:frL}
\end{figure}

 The critical value of the control-parameter for small $L$ is
given by
\begin{equation}
\Omega_c \simeq \frac{\kappa \pi^2}{4 a^2 \rho k L^2} 
+\frac{K+\lambda\alpha}{\rho k} +\frac{\alpha\gamma\xi_\perp L^2}
{\rho k a^2}
\quad . \label{eq:omc}
\end{equation}
For axoneme lengths below a characteristic value, $L<L_{\rm min}$, the
condition $\Omega_c\leq \pi^2$ is violated.  Therefore, oscillations
exist only for $L>L_{\rm min}$.  If $(K+\lambda\alpha)/\rho k\ll 1$,
$L_{min} \simeq\sqrt{\kappa / 4 a^2
\rho k}\simeq 1\mu$m for our choice of the parameters. 
The maximal frequency of oscillations occurs for $L=L_{\rm
min}$. Within the range of lengths between $1\mu m$ and $50 \mu m$, we
find frequencies $\omega_c/2\pi$ which vary between tens of Hz and
about $20 kHz$.  The parameters chosen in Fig. \ref{f:frL} lead to
frequencies above $150 Hz$. Lower frequencies are obtained for larger
values of the friction $\lambda$ or smaller chemical rate $\alpha$.

\subsection{Effect of external forces applied at the tail}

In the presence of an external force applied at the end $\bar\sigma
\neq 0$ and the filament pair is under tension. The eigenvalues
$\bar\chi_n$ depend on $\bar \sigma$, therefore the tension affects
shape and frequency of an unstable mode.  Inspection of
Eq. (\ref{eq:ep}) suggests that the tension $\sigma$ can be
interpreted as an additive contribution to the linear response
coefficient $\chi \rightarrow \chi + \sigma/a^2$. Taking into account
the boundary conditions complicates the situation, the
$\bar\sigma$-dependence of the eigenvalues is in general nontrivial.
However, for a filament with clamped head and a force applied at the
tail parallel to the tangent at the head (case D) the effect of
tension can be easily studied. In this particular case, the tension
leads to the same contribution to $\chi$ both in Eq. (\ref{eq:ep}) and
in the boundary conditions and the eigenvalues thus depend linearly on
$\sigma$:
\begin{equation}
\bar\chi_n(\bar\sigma,\bar\omega)=\bar\chi_n(0,\bar\omega)-\bar\sigma \quad .
\end{equation}
Since the tension corresponds to a change of the real part of $\chi$,
its presence has the same effect on filament motion as an increase of
the elastic modulus $K$ per unit length in Eq. (\ref{eq:linrep}) by
$\sigma/a^2$. Therefore, we can include the effects of tension on the
critical frequency in Eq. (\ref{eq:omegaapprox}) by replacing
$K\rightarrow K+\sigma/a^2$ which reveals that in those regimes where
the frequency depends on $K$, it will increase for increasing tension.
The critical value of the control parameter $\Omega_c$ is a function
of the tension, with Eq. (\ref{eq:omc}), we find
\begin{equation}
\Omega_c(\sigma)=\Omega_c(0)+\frac{1}{\rho k a^2} \sigma \label{eq:omsig}
\quad .
\end{equation}
For our choice of parameters, $\rho k a^2 \simeq 200$pN which
indicates that forces of the order of $10^2 pN$ should have a significant
effect on the bifurcation. Furthermore, Eq. (\ref{eq:omsig}) indicates
that the tension $\sigma$ can play the role of a second control
parameter for the bifurcation. Consider a tensionless filament
oscillating close to the bifurcation for $\Omega>\Omega_c(0)$.  If a
tension is applied, the critical value $\Omega_c(\sigma)$ increases
until the system reaches for a critical tension $\sigma_c$ a
bifurcation point with $\Omega_c(\sigma_c)=\Omega$ and the system
stops oscillating.

\section{Discussion}

In the previous sections, we have shown that many aspects of axonemal
eating can be described by a simple model based on the idea of local
sliding of microtubule doublets driven by molecular motors inside the
axoneme. This model represents a class of physical systems termed
internally driven filaments \cite{cama99} which have characteristic
properties that are closely linked to the geometric constraint that
couples global bending and local filament sliding.

Our work shows that axonemal beating can be studied both numerically
and analytically using a coarse-grained description which ignores many
details of the proteins involved and is based on effective material
properties such as bending rigidities, internal friction coefficient
and elastic modulus per unit length as well as the frequency dependent
linear response function of the active elements. Our main results are:
(i) The filament pair introduces a geometric coupling of the active
elements at different position along the filaments. This coupling is
suited for the generation of periodic beating motion by a general
self-organization mechanism of the system. (ii) The generation of
oscillations and propagating bending waves occurs via a Hopf
bifurcation for a critical value of a control-parameter such as the
ATP concentration. The patterns of motion generated close to this
bifurcation can be calculated without any knowledge of the internal
force generating mechanism if the oscillation frequency is
known. (iii) The frequency depends on chemical rates of the motors,
coefficients of internal elasticities and frictions, the solvent
viscosity and the microtubule rigidity.  Our calculations using a
simple two-state model for molecular motors suggest significant
variations of the oscillation frequency if the axonemal length is
varied. Short axonemes are predicted to be able to oscillate at
frequencies of several kHz.  This fact is particularly interesting in
the case of kinocilia which are located in the hair bundles of many
mechanosensitive cells that are involved in the detection of
sounds. If cilia are able to vibrate at high frequencies, this
suggests that the kinocilium could play as active role in sound
detection \cite{cama00}.  (iv) Forces applied to the axoneme can be an
important tool to study the mechanism of force generation. An external
force can play the role of a second control parameter for the
bifurcation which has a stabilizing effect for increasing forces.

These results have been obtained using several simplifying
assumptions.  Our model represents the solvent hydrodynamics by an
anisotropic local friction acting on the rod-like filament. This
approach ignores hydrodynamic interactions between different parts of
the filament which lead to logarithmic corrections.  These
hydrodynamic effects do not change the basic physics but can lead to
corrections to the numerical results. It is straightforward to
generalize our model to incorporate the effects of hydrodynamic
interaction, but this does not change our results qualitatively.  A
more serious simplification of our model is the restriction to a
two-dimensional system and to filament configurations which are
planar. This choice is motivated by the fact that observed bending
waves of flagella are planar in many cases \cite{brok91}. If filament
configurations are planar then our model applies and is complete.
However, this model cannot explain why motion is confined to a plane
and it misses all non-planar modes of beating. In order to address such
questions, a three dimensional generalizations have to be used. Such
generalizations introduce additional aspects.  In particular,
torsional deformations become relevant.  The local sliding
displacements depend in the three dimensional case both on bending and
torsional deformations and the full torsional dynamics has to be
accounted for in the dynamic equations.  Finally, we have restricted
most calculations to the limit of small deformations which corresponds
to filament shapes that are almost straight. This regime has several
important features. The filament dynamics can be studied by an
analytic approach by a systematic expansion in the deformation
amplitude.  This allows us to characterize linear and nonlinear terms
both for the filament dynamics as well as for the properties of the
active elements. Patterns of motion close to a Hopf bifurcation are
fully characterized by the linear terms of this expansion.  These
linear terms are given by the structure of the problem and their form
does not depend on molecular details of the system. For example, most
details of the operation of the molecular motors are unimportant for
the shape of the filament oscillations at the bifurcation, only the
linear response function of the active material plays a role. 

If filament beating with larger amplitude is of interest,
nonlinearities become relevant.  Nonlinearities arise due to nonlinear
geometric terms in the bending energy or via non-Hookian corrections
of the elasticity. Furthermore, nonlinearities in the force-generation
process of molecular motors exist.  All these nonlinearities determine
the large amplitude motion and could give rise to new types of
behavior such as additional dynamic instabilities. However, in
contrast to the linear terms, the form of the dominant nonlinearities
does depend on structural details of the axoneme. Therefore, an
analysis of large amplitude motion is difficult and knowledge of the
nature of the dominating nonlinearities is required. However, if no
new instabilities after the initial Hopf-bifurcation occur, the
principal effect of nonlinear terms is to fix the amplitude of
propagating waves. We therefore expect that propagating waves with
larger amplitude are in many cases well approximated by our
calculation to linear order.

Our work shows that propagating bending waves and oscillatory motion
of internally driven filaments can occur naturally by a simple
physical mechanism. Complex biochemical networks to control the system
are not required for wave propagation to occur. This suggests that the
basic axonemal structure intrinsically has the ability to oscillate.
Biochemical regulation systems are thus expected to control this
activity on a higher level but are not responsible for
oscillations. Our work shows that experimental studies of axonemal
beating close to a Hopf bifurcation would be very valuable.  Such
experiments could be e.g. performed by using demembranated flagella
and ATP concentrations not far from the level where beating sets
in. The observed behavior close to the bifurcation would give insight
in the self-organization at work.  Furthermore, externally applied
forces and manipulation of boundary conditions could be sensitive
tools to test some of the predictions of our work.

\acknowledgements
We thank A. Ajdari, M. Bornens, H. Delacroix, T. Duke, R. Everaers, K. Kruse,
A. Maggs, A. Parmeggiani, M. Piel, J. Prost and C. Wiggins
for useful discussions.

\appendix

\section{Functional derivative of the enthalpy}

The variation $\delta G$ of the enthalpy $G$ given
by Eq. (\ref{eq:G}) under variations $\delta{\bf r}$ of a shape ${\bf
r}(s)$ is
\begin{equation}
\delta G =\int_0^L \left[\left(\kappa C -a F \right) \delta C 
+ \Lambda \dot {\bf r} \cdot  \delta \dot {\bf r} \,\, \right]ds 
\label{eq:cp} \quad .
\end{equation}  
Note, that $\dot{\bf r}^2=1$ but $(\dot{\bf r}+\delta\dot{\bf
r})^2\neq 1$ in general.  Under such a variation, $\delta C = \delta
({\bf n} \cdot
\ddot{\bf r})= {\bf n} \cdot \delta\ddot{\bf r}$ since $ \ddot{\bf r} \cdot\delta{\bf
n}= C {\bf n} \cdot \delta{\bf n}=0$.  Therefore,
\begin{equation}
\delta G=\int_0^L \left[\left(\kappa C - a F \right) {\bf n} 
\cdot \delta \ddot {\bf r} + \Lambda {\bf t} \cdot \delta \dot {\bf r} 
\,\, \right]ds \quad .
\end{equation}
Two subsequent partial integrations lead to
\begin{eqnarray}
& & \delta G = 
\left[ \left(\kappa C -a F \right) {\bf n}\cdot \delta \dot {\bf r} \right]_0^L\label{eq:dG} \\ & & +\left[ \left( -(\kappa \dot C  -a f) {\bf n}+(\kappa C^2  -a C F
+\Lambda) {\bf t} \right) \cdot \delta {\bf r} \right]_0^L \nonumber \\  & & + 
\int_0^L \partial_s \left( (\kappa \dot C  -a f) {\bf n}-
(\kappa C^2  -a C F +\Lambda ) {\bf t}\right) \cdot \delta {\bf r} \,\,ds \quad . \nonumber
\end{eqnarray}
The functional derivative $\delta G/\delta {\bf r}$ given by
Eq. (\ref{eq:dGdr}) can be read off from the integrand, the boundary
terms provide expressions for forces and torques at the ends given by
Eqs.  (\ref{eq:bcla}) and (\ref{eq:bcle}).

\section{Small deformations}

Inserting the expansions (\ref{eq:exqpsi}) in the differential
Eq. (\ref{eq:li}) for the tension profile leads at each order in
$\epsilon$ to a separate equation. Up to second order we find
\begin{eqnarray}
\ddot \tau_0 &=& 0 \label{eq:tpqo} \\
\ddot \tau_1 &=& 0 \nonumber \\ 
\ddot \tau_2 &=& \partial_s(\dot\psi_1(a f_1-\kappa \ddot \psi_1) )
+ \frac{\xi_{\parallel}}{\xi_{\perp}} \dot \psi_1( -\kappa \stackrel{\mbox{...}}{\psi_1}+ a \dot f_1 + \tau_0 \dot\psi_1 ) \quad . \nonumber
\end{eqnarray}
From the boundary conditions, it follows that $\tau_0=\sigma$ is
constant and $\tau_1=0$.  Repeating the same procedure for the dynamic
Eq. (\ref{eq:emt}) using $\tau_1=0$ and $\tau_0=\sigma$, we obtain
\begin{eqnarray}
\xi_{\perp} \partial_t \psi_1 &=& -\kappa\stackrel{\mbox{....}}{\psi_1} + 
a \ddot f_1 +\sigma \ddot \psi_1 \nonumber \\
\xi_{\perp} \partial_t \psi_2 &=& - \kappa \stackrel{\mbox{....}} {\psi_2}
\label{eq:emtsd}
\end{eqnarray} 
The Equation for $\psi_2$ is independent of $f_1$ and only describes
transient behavior which depends on initial conditions. After long
times and for limit cycle motion, $\psi_2=0$.  The transverse and
longitudinal displacements $h$, $u$ and the velocity $\bar v$ are
obtained perturbatively as
\begin{eqnarray}
h &=&\epsilon h_1+\epsilon^2 h_2+O(\epsilon^3) \nonumber\\
u &=&\epsilon u_1+\epsilon^2 u_2+O(\epsilon^3) \nonumber\\
\bar v &=&\epsilon \bar v_1+\epsilon^2 \bar v_2+O(\epsilon^3)  
 \label{eq:sdqov}   \quad .
\end{eqnarray}
Using Eqns. (\ref{eq:exqpsi}) and (\ref{eq:sdfromtv}) we find
$u_1(s)=u_1(0)$, $h_2(s)=h_2(0)$ and
\begin{eqnarray}
h_1(s) &=& h_1(0) + \int_0^s \psi_1(s') ds' \label{eq:exheady}\\
u_2(s) &=& u_2(0) -\frac{1}{2} \int_0^s \psi_1(s')^2 ds'
\label{eq:exheadx} \quad .
\end{eqnarray}
The dynamics of $h(s)$ and $u(s)$ to second order in $\epsilon$ is
therefore determined by $\psi_1(s,t)$ and the motion $\partial_t{\bf
r}$ for $s=0$.  For the latter, we find from Eq. (\ref{eq:gem2}) $\bar
v_1=0$, $\partial_t u_1(0)=0$, $\partial_t h_2(0)=0$ and
\begin{eqnarray} 
\xi_{\perp} \partial_t h_1(0) &=& -\kappa \stackrel{\mbox{....}}{h_1}(0)+
\tau_0 \ddot h_1(0) +a\dot f_1(0) \label{eq:bch1}\\
\xi_{\parallel}(\bar v_2+\partial_t u_2(0)) &=&-\xi_{\parallel}\psi_1(0)\partial_t h_1(0) +
 \kappa \dot \psi_1(0) \ddot
 \psi_1(0)\nonumber \\
 &-& a \dot \psi_1(0) f_1(0) + \dot \tau_2(0) \quad . \label{eq:bcuv2} 
\end{eqnarray}
For the lateral motion, we obtain with Eq. (\ref{eq:emtsd}) and
(\ref{eq:exheady}),
\begin{eqnarray}
\xi_{\perp}(\partial_t h_1(s) - \partial_t h_1(0)) &=&
\left[ -\kappa\stackrel{\mbox{....}}{h_1}+\sigma \ddot h_1
+a \dot f_1 \right]_0^s
\label{eq:Ymot}
\end{eqnarray}
Using Eq. (\ref{eq:bch1}), we obtain 
\begin{equation}
\xi_{\perp} \partial_t h_1= 
-\kappa \stackrel{\mbox{....}}{h_1}+\sigma \ddot h_1 +a\dot f_1 \quad , 
\label{eq:ldqoh}
\end{equation}
which is the result given in Eq. (\ref{eq:emld}).  In order to
calculate $\tau_2(s)$ and $\bar v_2$ we integrate
Eq. (\ref{eq:tpqo}). Together with Eq. (\ref{eq:ldqoh}),
\begin{eqnarray}
\dot \tau_2 (s) &=& \dot \tau_2 (0)+\left[ \dot \psi_1 
(a f_1 - \kappa \ddot \psi_1) \right]_0^s \nonumber \\
&+& \xi_{\parallel} \int_0^s \dot \psi_1 \partial_t h_1 ds' \quad .
\end{eqnarray}

After elimination of $\dot \tau_2(0)$ via
Eq. (\ref{eq:bcuv2}) we find the tension profile
\begin{eqnarray}
\tau_2(s) & = & \tau_2(0)+ s \xi_{\parallel} (\bar v_2+\partial_t u_2(0)) \nonumber \\
& + & \int _0^s ds '\left[\dot \psi_1 (a f_1 - \kappa \ddot \psi_1) +\xi_{\parallel} 
\psi_1 \partial_t h_1\right ]  \nonumber \\
&-& \xi_{\parallel} \int_0^s ds'\int_0^{s'} ds''  \psi_1 
\partial_t \psi_1 
\label{eq:dstau2} \quad .
\end{eqnarray}
The tension at the head $\tau_2(0)$, as well as $\bar v_2 + \partial_t
u_2(0)$ are fixed using the boundary conditions. In cases A,B and D
where the head is fixed $\bar v_2 + \partial_t u_2(0)=0$. In this case,
$\tau_2(0)$ follows from Eq. (\ref{eq:dstau2}) at $s=L$. For
boundary condition C, the velocity of the head is determined from
the condition $\tau_2(L)=0$ together with the boundary
conditions $\tau_2(0)=-\psi_1(0)(\kappa \ddot
\psi_1(0)-af_1(0))+\zeta(\bar v_2+\partial_t u_2(0))$ and 
$\kappa \stackrel{\mbox{...}}{h_1}(L)=af_1(L)$ which 
leads to
\begin{eqnarray} 
\bar v_2+\partial_t u_2(0) &=& \frac{\xi_{\perp}-\xi_{\parallel}}{\zeta+\xi_{\parallel}L} 
\int_0^L \dot h_1 
\partial_t h_1 ds \nonumber \\ 
&+&\frac{\xi_{\parallel}}{\zeta+\xi_{\parallel}L} 
\int_0^L ds
\int_0^s ds'\frac{1}{2} \partial_t ({\dot h_1}^2) 
\label{eq:dtX20}  \quad .
\end{eqnarray}
Averaging this equation  for periodic motion
$h_1(s,t)=H(s)\cos(\omega t-\phi(s))$ leads to Eq. (\ref{eq:noloadavve}).

\section{Linear and nonlinear response functions of the two-state model}

We introduce the Fourier modes of the distribution function
$P_1(\xi,t)=\sum_k P_1(\xi,k) e^{ik\omega t}$. The dynamic
equation (\ref{eq:twost}) of the two-state model can then be
expressed as
\begin{equation}
P_1(\xi,k)=\delta_{k,0}R(\xi)-i\omega \sum_{lm}
\frac{l \delta_{k,l+m}}{\alpha+i\omega k} \Delta_l\partial_\xi P_1(\xi,m)
\label{eq:p1k}
\end{equation}
Inserting the Ansatz
\begin{equation}
P_1(\xi,k)=R\delta_{k,0}+P^{(1)}_{kl} \Delta_l + P^{(2)}_{klm}\Delta_l
\Delta_m + O(\Delta^3)
\end{equation}
into Eq. (\ref{eq:p1k}), one obtains a recursion relation 
\begin{equation}
P^{(n)}(\xi)_{k,k_1,..,k_n}=-i\omega \sum_l\frac{l \delta_{k,k_n+l}}
{\alpha+i\omega k}\partial_\xi P^{(n-1)}_{l,k_1,..,k_{n-1}} \quad .
\end{equation}
the knowledge of the functions $P^{(n)}_{k,k_1,..,k_n}$ allows one
to calculate the coefficients of the expansion (\ref{eq:fnexp}).
In particular, with $P^{(0)}=R$, one obtains the result
(\ref{eq:F1}) and
\begin{equation}
F^{(n)}_{k,k_1,..,k_n}=\rho \int_0^l d\xi P^{(n)}_{k,k_1,..,k_n}(\xi)
\partial_\xi\Delta W
\end{equation}

\section{Eigenvalues and Eigenmodes near a Hopf bifurcation}

Nontrivial solutions to Eq. (\ref{eq:ep}) can be found using the
Ansatz $\tilde h= A e^{qs/L}$ with $q$ being solution to
\begin{equation}
q^4 - (\bar \sigma + \bar \chi) q^2 +i \bar \omega =0 \label{eq:ki}
\end{equation}
Here, $\bar \sigma \equiv \sigma L^2 /\kappa$, $\bar \omega \equiv
\omega \xi_{\perp} L^4/\kappa$, and $\bar \chi \equiv \chi a^2 L^2
/\kappa$ are a dimensionless tension, frequency and linear
response coefficient, respectively.  Eq. (\ref{eq:ki}) has four
complex solutions $q_i$, $i=1,..,4$.  Therefore,
\begin{equation}
\tilde h = \sum_{i=1}^4 A_i e^{q_i s/L} \quad .
\end{equation}
In order to determine the amplitudes $A_i$ the four boundary
conditions have to be used.  Since Eq. (\ref{eq:ep}) is linear and
homogeneous, the equations for the coefficients $A_i$ have the form
\begin{equation}
\sum_{i=1}^4 A_i M_{ij}=0 \quad ,  \label{eq:epM}
\end{equation}
where $M_{ij}$ is a $4\times 4$ matrix which depends on $\bar\chi$,
$\bar\sigma$ and $\bar \omega$. Nontrivial solutions exist only for
those values $\bar\chi=\bar\chi_n$ for which $\det M_{ij}=0$. The
corresponding eigenmode is the solution for $A_i$ of
Eq. (\ref{eq:epM}). These eigenvalues and eigenfunctions can be
determined by numerically obtaining solutions for $\det M_{ij}=0$.  In
the limit of small $\bar \omega$, analytic
expressions can be obtained by inserting the ansatz $\bar \chi = \bar
\chi^{(0)} + \bar \chi^{(1)}
\bar \omega +O({\bar \omega}^2)$ for the eigenvalue which leads to an
expansion of $\det M_{ij}$ in powers of ${\bar \omega}^{1/2}$.  This
procedure leads for $\bar\sigma=0$ to $\bar \chi^{(0)}_n=-[(n-1)\pi + 
\pi/2]^2$ independent of boundary conditions. The linear coefficient
$\bar\chi^{(1)}_1=-i\gamma$ depends on boundary conditions. For boundary
conditions  A
\begin{equation}
\gamma=\frac{12\pi^3-32\pi^2+1728\pi-4608}{\pi^5+144\pi^3} 
\simeq 0.184  \quad ,
\end{equation}
while for case B,
\begin{equation}
\gamma= \frac{192}{\pi^7} \left( \frac{\pi^5}{32}+\frac{\pi^4}{12}-\pi^3+2\pi^2-2\pi \right) \simeq 0.008 \quad .
\end{equation}

\end{document}